\documentclass[aps,nofootinbib,superscriptaddress,twocolumn]{revtex4}

\usepackage{float}
\usepackage{graphicx}  
\usepackage{subfig}
\usepackage{amsmath}   
\usepackage{amssymb}   
\usepackage{bm} 
\usepackage{dcolumn}
\usepackage{color}
\usepackage{mathrsfs}
\usepackage{amsfonts}
\usepackage{varioref}
\usepackage{textcomp}
\RequirePackage[colorlinks,citecolor=blue,urlcolor=magenta,linkcolor=blue]{hyperref}
\allowdisplaybreaks

\labelformat{section}{Section #1} 
\labelformat{subsection}{Section #1} 
\labelformat{subsubsection}{Section #1}
\labelformat{subsubsubsection}{Section #1}
\labelformat{equation}{Eq.~(#1)} 
\labelformat{figure}{Fig.~#1} 
\labelformat{subfigure}{Fig.~\thefigure#1} 
\labelformat{table}{Tab.~#1} 
\begin{document}

\title{Dynamical Love for Area Quantized Black holes}

\author{Sreejith Nair}
\email{sreejithnair@iitgn.ac.in}
\affiliation{Indian Institute of Technology, Gandhinagar, Gujarat-382355, India}

\author{Sumanta Chakraborty}
\email{tpsc@iacs.res.in}
\affiliation{School of Physical Sciences, Indian Association for the Cultivation of Science, Kolkata-700032, India}

\author{Sudipta Sarkar}
\email{sudiptas@iitgn.ac.in}
\affiliation{Indian Institute of Technology, Gandhinagar, Gujarat-382355, India}

  
\begin{abstract}
We study the tidal deformability of horizonless exotic compact objects, with implications for area quantized black holes. Since, any such horizonless compact object, including area quantized black hole, possesses a non-zero reflectivity, it follows that they inherit a nonzero tidal Love number, which varies as a function of the perturbing frequency. The reflectivity of the horizon for an area quantized black hole has a distinct shape and unmistakable features, which are also present in the frequency-dependent tidal Love numbers, and therefore is a smoking gun test for such quantum black holes. The existence of these features also promotes the dynamical Love number to be a crucial observational tool to not only distinguish area quantized black holes from other models of horizonless compact objects, but also to differentiate between black holes with distinct schemes of area quantization. We further discuss implications of the same for the present and the future gravitational wave detectors.
\end{abstract}
\maketitle
\section{Introduction}\label{section:intro}

Gravitational wave measurements \cite{LIGOScientific:2016aoc, LIGOScientific:2016lio} have provided us an opportunity to probe the near-horizon physics to an unprecedented level and study the nature of the compact objects sourcing these waves. In particular, we hope to test the black hole hypothesis, which asserts the existence of a one-way null surface, namely the event horizon. 

As an alternative to the black hole hypothesis, various models of horizon-less compact objects are known in the literature, like gravastars \cite{Mazur:2001fv,Visser:2003ge}, boson stars \cite{Jetzer:1991jr,Schunck:2003kk,Liebling:2012fv}, and quantum black holes \cite{Bekenstein:1974jk,Bekenstein:1995ju,Ashtekar:1997yu,Ashtekar:2000eq}, all of which share the feature of gravitational perturbations not being completely absorbed by the horizon. Therefore, unlike a classical black hole, they possess a non-zero reflectivity. The lack of complete absorption can also be a consequence of possible quantum modifications of the theory of gravity near the horizon. As expected, the search for observational signatures of such non-zero reflectivity is a domain of active research \cite{Hod:1998vk,Maggiore:2007nq,Macedo:2013qea,Laghi:2020rgl,Cardoso:2017cfl,Agullo:2020hxe,Chakraborty:2022zlq}. Here we shall demonstrate the consequences of such a non-trivial reflectivity on the tidal deformability of the compact object, measured by the so-called Love numbers.

The non-zero reflectivity may result from a horizonless compact object with a surface reflecting the perturbation outwards. Then, the perturbation in each frequency will be reflected completely or partially. The details of the reflected wave will depend on the internal physics of the system. One of the possibilities being, an area-quantized black hole \cite{Bekenstein:1974jk}, which absorbs only at selected frequencies, similar to an atomic system. Further refinement of the area quantized black holes can be found in the pioneering work by Bekenstein and Mukhanov \cite{Bekenstein:1995ju}. Then the black hole can only undergo transitions to discrete mass/energy levels as it interacts with external perturbations, similar to the atomic transitions. Therefore, one expects non-trivial physics to emerge from the quantum nature of the black hole, which will manifest through a discrete reflectivity on the surface $r=r_{+}+\epsilon$, where $r_{+}$ is the location of the horizon and $\epsilon$ is expected to be $\mathcal{O}(l_{\rm pl})$. Among various possible models of such quantum black holes,  describes the area-quantized black holes. The area quantization has many interesting consequences on the emission spectrum of black holes, namely on the profile of the Hawking radiation \cite{Bekenstein:1995ju,Chakraborty:2017opo,Chakraborty:2017pmn,Chakraborty:2016fye,Lochan:2015bha} as well as on the quasi-normal mode spectrum \cite{Foit:2016uxn, Cardoso:2019apo, Agullo:2020hxe, Datta:2020gem, Datta:2021row, Chakravarti:2021jbv, Chakravarti:2021clm}. Though it has its origins as a phenomenological model, area quantized black holes have also been constructed from a first principle approach from certain calculations in Loop Quantum Gravity \cite{FernandoBarbero:2009ai}.

Any such modification of the boundary condition at the horizon due to non-zero reflectivity is expected to affect the system's response to external tidal perturbations. Hence determining the tidal deformability of such an exotic compact object (ECO), characterized by the tidal Love numbers, is of significant interest. Since the tidal deformation of the classical black hole solutions of general relativity vanishes identically \cite{Binnington:2009bb,Damour:2009vw,Kol:2011vg,Pani:2015hfa,Landry:2015zfa,LeTiec:2020spy,Chia:2021}, one would like to know whether systems like quantum black holes can be deformed by a perturbing tidal field which may manifest as a non-zero tidal Love number for such systems. Furthermore, it will be of interest to determine if these results depend on the details of the nature of quantization.

In this work, we report the existence of a non-zero frequency-dependent (dynamical) tidal Love number for any exotic compact horizon-less object, generalizing earlier results in the literature \cite{Cardoso:2017cfl}. In particular, if the reflectivity is \textit{non-zero}, we have the dynamical, i.e., frequency-dependent tidal love numbers to be \textit{non-vanishing} as well. Further, we discuss its observational signature for specific cases, e.g., in the context of area quantized black holes. Our results provide a new tool to study the near-horizon physics using gravitational-wave observations, particularly the nature of compact objects, in the dynamical situations. 

The work is organized as follows: In \ref{area_quant}, we provide a brief review of various models of the area-quantized black holes, as a proxy for exotic compact objects. Then, in \ref{dyn_tln}, we provide the definition of the tidal Love number in the Newtonian context and subsequently generalized it to the relativistic context in \ref{dyn_tln_gr}. Finally, the reflectivity of an area-quantized black hole has been presented in \ref{refquant}, using which the computation of the tidal Love number has been performed in \ref{imp_gw} to study the observability of these dynamical tidal Love numbers in the context of gravitational wave measurements. Finally, we conclude with a discussion of the results. 

\emph{Notations and Conventions:} Throughout the paper, we have used the mostly positive signature convention, i.e., the flat spacetime metric in the Cartesian coordinates look like: $\eta_{\mu \nu}=\textrm{diag.}\left(-1,+1,+1,+1\right)$. We have also set the fundamental constants $G$, $\hbar$, and $c$ to unity.  
\section{Area quantized black holes as exotic compact objects}\label{area_quant}

Before we go ahead with the discussion on the consequence of the nonzero reflectivity of ECOs on the tidal love numbers, let us first discuss a few of the models of ECOs, with an emphasis on the area quantized black holes. 

We may start with the gravastar, which is an ECO having an interior de-Sitter spacetime matched to a Schwarzschild near the horizon \cite{Mazur:2001fv,Visser:2003ge}. This system has no horizon but a matching timelike surface with nonzero reflectivity, which depends on the details of the matching. Next, we have the boson star \cite{Jetzer:1991jr,Schunck:2003kk,Liebling:2012fv}, which arises as solutions to theories with complex scalar fields like the Einstein-Klein-Gordon theory. In these objects, most of the scalar field is contained within a so-called effective radius. Again, the absence of a horizon means that these objects can be treated as having a non-vanishing reflectivity at some $r$, larger than the effective radius.

There also exist objects called quantum black holes \cite{Bekenstein:1974jk,Bekenstein:1995ju,Ashtekar:1997yu,Ashtekar:2000eq}, which are spacetimes with a discrete nonzero reflectivity enforced at $r=r_{+}+\epsilon$ ($\epsilon\sim l_{\rm pl}$) as a result of some quantum theory of gravity. Here, $r_{+}$ is the location of the outer event horizon of the black hole spacetime. The quantum black holes predicted by these theories of gravity have a quantization rule, which depends on the specifics of the problem. Instead of getting involved in the intricate physics of quantum gravity, we may follow the route taken by Bekenstein and Mukhanov in \cite{Bekenstein:1974jk, Bekenstein:1995ju}, where they considered the black hole area to be quantized in linear steps. This can be motivated by considering the black hole area to be an excited state of a bunch of harmonic oscillators generating the area, such that (restoring $c$, $G$, and $\hbar$ for the moment),
\begin{equation}
A=\alpha \ell_{\rm p}^{2}N~.
\end{equation}
Here, $N$ is an integer, $\ell_{\rm p}=\sqrt{\hbar G/c^{2}}$ and $\alpha$ depends on the specifics of the quantum gravity. 

In Bekenstein's original work \cite{Bekenstein:1974jk}, $\alpha$ is chosen to be $8\pi$. Whereas, using arguments based on \textit{corresponding principle}, a value $\alpha = 4 \ln 3$ is proposed in \cite{Hod:1998vk}. Later, in \cite{Maggiore:2007nq}, considering both real and imaginary values of the black hole quasinormal modes, $\alpha = 8\pi$ is put forward. Also, models involving the counting of microstates suggested $\alpha=4\log k$, $k$ being an integer;  with the lowest value being $4\log2$ \cite{Hod:1998vk}. In what follows, we will keep $\alpha$ as an arbitrary $\mathcal{O}(1)$ constant and demonstrate our result for various possible choices of $\alpha$.

As a consequence of quantization, the black hole cannot absorb all possible frequencies but will do so in discrete steps. For a rotating area quantized black hole, the frequency associated with the transition in the area from $N\rightarrow (N+n)$, corresponds to \cite{Agullo:2020hxe},  
\begin{equation}\label{uniform_area}
\omega_{N,n}=\left(\frac{\alpha \kappa}{8\pi}\right)n+2\Omega_{\rm h}+\mathcal{O}\left(N^{-1}\right)~.
\end{equation}
Here the angular momentum is also quantized as $J=\sqrt{j(j+1)}\hbar\simeq j\hbar$ (assuming $j\gg 1$) and the above frequency corresponds to a transition in the angular momentum, as $j\rightarrow j+2$. Also, in the above expression, $\kappa$ is the surface gravity, and $\Omega_{\rm h}$ is the angular velocity of the horizon for the Kerr spacetime. Note that the frequencies depend explicitly on the details of the quantizations through the parameter $\alpha$ and on the background spacetime through $\kappa$ and $\Omega_{\rm h}$. 

Since most of the quantum gravity models involve sub-leading corrections to the entropy of the black hole, which will be translated in the quantized area spectrum as well, we expect it to be reflected in the area quantized black hole absorption spectrum \cite{Kothawala:2008in, Chakravarti:2021jbv}, 
\begin{equation}
A=\alpha \ell_{\rm p}^{2}N\left(1+CN^{\nu}\right)~.
\end{equation}
Here, $\alpha$, $C$, and $\nu$ are the parameters arising from the underlying quantum theory of gravity. This would imply that, when a transition from $N$ to $(N+n)$ area level and from $j$ to $(j+2)$ angular momentum level happens, the black hole absorbs the following frequency \cite{Kothawala:2008in,Chakravarti:2021jbv},
\begin{equation}\label{subleading_area}
\omega_{N,n}=\left(\frac{\alpha \kappa}{8\pi}\right)\left[1+C(1+\nu)N^{\nu}\right]n+2\Omega_{\rm h}+\mathcal{O}\left(N^{-1}\right)~.
\end{equation}
Note that in the limit of $C\rightarrow 0$, i.e., when there are no sub-leading corrections to the black hole entropy, we recover \ref{uniform_area}, as expected. Thus, the black hole only absorbs the incoming radiation in the above frequencies, which depends on the details of the quantization scheme and the nature of the background spacetime.
\section{Dynamical Tidal Love numbers in Newtonian Gravity}\label{dyn_tln}

The tidal Love number for any compact object is essentially a measure of the deformation of the respective object caused by an external tidal field $\mathcal{E}$, whose spherical harmonic decomposition is governed by the components $\mathcal{E}_{\ell m}$ \cite{Poisson(Book):2014,Zhang:1986cpa}. These Love numbers are a faithful probe of the internal structure of the tidally deformed compact objects. In general, the tidal Love number $\lambda_{\ell m}$ of a compact object is given by the ratio of the induced moment $Q_{\ell m}$ associated with the $(\ell,m)$ mode of the spherical harmonic decomposition, with the external tidal field itself. From this definition, it is evident that $\lambda_{\ell m}$ has the dimension of $(\textrm{mass})^{2\ell+1}$, in units where the speed of light and Newton's constant have been chosen to be unity. It is often advantageous to define the dimensionless tidal Love number $k_{\ell m}$ through the following relation: $k_{\ell m}\sim (\lambda_{\ell m}/R^{2\ell+1})$, where $R$ is the length scale of the compact object. Moreover, the tidal Love numbers, in general, can be divided into electric and magnetic parts, depending on their parity. We would like to mention that even though the tidal Love numbers can be derived for all the spherical harmonics $(\ell,m)$. The most dominant contribution will come from the electric part and for quadrupolar deformation, corresponding to the choice $\ell=2$ and is the one we will also concentrate on. In the general case, the change in the multipole moment $Q_{\ell m}$ due to an external time-varying tidal field $\mathcal{E}_{\ell m}$ is expressed as \cite{Poisson(Book):2014},
\begin{equation}\label{deltaQ}
\delta Q_{\ell m}=-\frac{(\ell-2)!}{(2\ell-1)!!}R^{2\ell+1}(2k_{\ell m}\mathcal{E}_{\ell m}-\tau_{0}\nu_{\ell m}\Dot{\mathcal{E}}_{\ell m})~,
\end{equation}
where, $R$ is the length scale of the compact object and $k_{\ell m}$ are the tidal Love numbers, while $\tau_{0}$ and $\nu_{\ell m}$'s are related to the tidal absorption of the deformed object. The tidal Love numbers $k_{\ell m}$ can also be computed from the change in the gravitational potential due to the deformation induced by a time dependent $\mathcal{E}_{\ell m}$ and hence takes the form \cite{Poisson(Book):2014},
\begin{align}
\delta U&=\sum_{\ell m}\mathcal{U}_{\ell m}Y_{\ell m}(\theta,\phi)~;
\nonumber
\\
\mathcal{U}_{\ell m}&=\frac{r^{\ell}\mathcal{E}_{\ell m}}{\ell(\ell-1)}\left[1+\left(2k_{\ell m}+i\omega \tau_{0}\nu_{\ell m}\right)\left(\frac{R}{r}\right)^{2\ell+1}\right]~.
\label{deltaU}
\end{align}
Here, we have moved from time-domain to the frequency-domain and hence the time derivative of the tidal field in \ref{deltaQ} leads to a factor of $-i\omega$ in the above expression. Thus, if we can read off the perturbation to the gravitational potential, then the coefficient of the decaying term $r^{-2\ell-1}$ captures the tidal effects, with the real part giving rise to tidal Love number and the imaginary part gets related to the tidal dissipation. In the next section, we will see how this notion may be extended to relativistic objects.

\section{Dynamical Tidal Love numbers}\label{dyn_tln_gr}

In order to generalize the notion of tidal effects in fully relativistic context, we need to start from a coordinate invariant formulation, or in other words we need a gauge invariant formulation of the tidal Love number. In this context, it is useful to consider the Weyl scalars, since these faithfully captures the Newtonian potential in appropriate limits, while remaining gauge invariant. Since the tidal Love numbers are derived by constructing an asymptotic expansion of the deformed potential sourced by an external tidal field near infinity, we may use the Weyl scalar $\psi_4$ for our purpose \cite{Newman:1961qr,Szekeres:1965ux}. The dynamics of the Weyl scalar $\psi_4$, associated with the perturbations of the Kerr geometry of mass $M$ and angular momentum $a$, can be separated into a radial and an angular part, as \cite{Teukolsky:1972my,Teukolsky:1974yv,Teukolsky:1973ha},
\begin{align}\label{psi4}
\big(r-ia\cos\theta &\big)^4\psi_{4,\ell m}(v,r,\theta,\widetilde{\phi};\omega)
\nonumber
\\
&=e^{-i\omega v}e^{im\widetilde{\phi}}S_{\ell m}(\theta,\omega)\mathcal{R}_{\ell m}(r,\omega)~,
\end{align}
where, $(v,r,\theta,\widetilde{\phi})$ corresponds to the ingoing null coordinates and the Weyl scalar $\psi_{4}$ can be obtained by integrating the above expression over $\omega$ and then summing over all possible choices of $\ell$ and $m$. The angular part $S_{\ell m}$ can be solved to yield spin-weighted spheroidal harmonics, while the radial part satisfies the following equation \cite{Teukolsky:1972my,Teukolsky:1974yv,Teukolsky:1973ha} 
\begin{align}
\Big[\Delta \frac{d^{2}}{d r^{2}}&+2[(s+1)(r-M)-iK]\frac{d}{d r}
\nonumber
\\
&+2(2 s-1) i \omega r-\lambda \Big]\mathcal{R}_{\ell m}(r,\omega)=0~,
\end{align}
with, $K=(r^2+a^2)\omega-am$, $\lambda=E-2a m \omega + a^2\omega^2-s(s+1)$, $s=-2$ and $\Delta = r^2 - 2 a M + a^2$, obtained using the Kinnersley’s tetrad. To get an analytic handle to the above equation, we may consider the low-frequency approximation ($M\omega\ll 1$), such that the radial part $\mathcal{R}_{\ell m}(r,\omega)$ of the gravitational perturbation can be described by the following differential equation \cite{Chia:2021},
\begin{align}
\label{radialdif}
&\frac{d^{2}\mathcal{R}_{\ell m}(\xi,\omega)}{d \xi^{2}}+\left[\frac{\left(2 i p-1\right)}{\xi}-\frac{\left(2 i p+1\right)}{(\xi+1)}\right] \frac{d \mathcal{R}_{\ell m}(\xi,\omega)}{d \xi}
\nonumber
\\
&+\left[\frac{4 i p}{(\xi+1)^{2}}-\frac{4 i p}{\xi^{2}}-\frac{[\ell(\ell+1)-2]}{\xi(\xi+1)}\right] \mathcal{R}_{\ell m}(\xi,\omega)=0~.
\end{align}
The dimensionless parameters $\xi$ and $p$ are defined as, $\xi \equiv (r-r_+)/(r_+-r_-)$ and $p\equiv (am-2M\omega r_+)/(r_+-r_-)$. The above equation has two linearly independent solutions in terms of the Hypergeometric functions, such that the most general solution is given by,
\begin{align}\label{radial}
\mathcal{R}_{\ell m}(\xi,\omega)=A\mathcal{R}^{\rm in}_{\ell m}(\xi,\omega)+B\mathcal{R}^{\rm out}_{\ell m}(\xi,\omega)~,
\end{align}
where, the two linearly independent solutions $\mathcal{R}^{\rm in}_{\ell m}(\xi,\omega)$ and $\mathcal{R}^{\rm out}_{\ell m}(\xi,\omega)$ take the following forms,
\begin{align}
\mathcal{R}^{\rm in}_{\ell m}(\xi,\omega)&=(1+\xi)^{2} \xi^{2}~{}_2F_{1}\left[2-\ell, 3+\ell; 3+2ip;-\xi \right]~,
\\
\mathcal{R}^{\rm out}_{\ell m}(\xi,\omega)&=(1+\xi)^{2} \xi^{-2ip}
\nonumber
\\
&\hskip 0.1 cm \times ~{ }_{2} F_{1}\left[-\ell-2 i p, 1+\ell-2 i p;-1-2 i p ;-\xi \right]~.
\end{align}
The above functions are referred to as the ingoing and the outgoing solutions, since in the near horizon limit, $\mathcal{R}^{\rm in}\sim (r-r_+)^2$ and $\mathcal{R}^{\rm out}\sim (r-r_+)^{-2ip}$, providing exactly the ingoing and the outgoing boundary conditions for $\psi_4$, respectively \cite{Teukolsky:1974yv}. For a classical black hole spacetime, there should not be any outgoing solution near the horizon, and hence the coefficient $B$ must vanish in \ref{radial}, and $\psi_{4}$ is governed by $\mathcal{R}_{\rm in}(\xi)$ alone.

Since we are interested in horizon-less ECOs, we cannot set $B=0$, and we must have contributions from both $\mathcal{R}^{\rm in}$ and $\mathcal{R}^{\rm out}$ in the solution of the radial perturbation equation. The nonzero reflectivity of the compact object is then defined as $\mathcal{R}\equiv B/A$. It will contain information about the specific properties of the ECO. We emphasize that the classical geometry remains valid till the reflective surface, which is generically considered to be located very close to the horizon, and all possible violations from classical black hole paradigm is encoded within this reflectivity $\mathcal{R}$. In order to find the tidal Love number, exact form of the reflectivity would be necessary, and that will distinguish between various models of ECO.

To arrive at the expression for the tidal Love number, we first expand the general solution of the radial function $\mathcal{R}_{\ell m}(\xi,\omega)$ in the tidal region \cite{Chia:2021}, characterized by the condition $\xi\gg1$, or, equivalently $(r/r_{+})\gg 1$, leading to
\begin{widetext}
\begin{equation}
    \begin{aligned}
        \mathcal{R}_{\ell m}&\simeq \frac{\Gamma(-2 \ell-1) \Gamma\left(2 i p+3\right)}{\Gamma(2-\ell) \Gamma(2 i p-\ell)} \xi^{1-\ell}\bigg[1+O\left(\frac{1}{\xi}\right)\bigg]
        +\frac{\Gamma(2 \ell+1) \Gamma\left(2 i p+3\right)}{\Gamma(\ell+3) \Gamma\left(\ell+1+2 i p\right)} \xi^{\ell+2}\bigg[1+O\left(\frac{1}{\xi}\right)\bigg]  \\
        \\
        &+ \mathcal{R}\Bigg\{ \frac{\Gamma(-2 \ell-1) \Gamma\left(-2 i p-1\right)}{\Gamma(-\ell-2) \Gamma\left(-\ell-2 i p\right)} \xi^{1-\ell}\bigg[1+O\left(\frac{1}{\xi}\right)\bigg]
        +\frac{\Gamma(2 \ell+1) \Gamma\left(-2 i p-1\right)}{\Gamma(\ell-1) \Gamma\left(\ell-2 i p+1\right)} \xi^{\ell+2}\bigg[1+O\left(\frac{1}{\xi}\right)\bigg]\Bigg\}~.\\
    \end{aligned}
\end{equation}
\end{widetext}
In the above equation, it should be understood that the term within the curly brackets arises because of the absence of a horizon. The black hole limit is obtained by simply setting $\mathcal{R}=0$. Using the above behaviour of $\mathcal{R}_{\ell m}(\xi,\omega)$ we may compute $\psi_{4,\ell m}(\omega)$ in the tidal region using \ref{psi4}, and we will get,
\begin{equation}
\begin{aligned}
\psi_{4,\ell m}(\omega) &\sim r^{\ell-2}\Big\{\left[1+\mathcal{O}(r^{-1})\right]\\
&+\left(\frac{R}{r}\right)^{2\ell+1}\left(2k_{\ell m}+i\omega\tau_{0}\nu_{\ell m}\right)[1+\mathcal{O}(r^{-1})]\Big\}~.
\end{aligned} 
\end{equation}
The first term in the above expansion is the growing term associated with the tidal source, while the second term encodes the response of the compact object. Here, once again, $k_{\ell m}$ is the dimensionless tidal Love number and $\nu_{\ell m}$ corresponds to the tidal dissipation \cite{Poisson(Book):2014}. The characteristic length scale associated with the object can be taken to be $R=r_{+}-r_{-}$, though for a compact object none of these radii have any physical significance. In the case of a black hole, it was demonstrated in \cite{Chia:2021}, that $k_{\ell m}$ identically vanishes, however, $\nu_{\ell m}$ is non-zero. In the present scenario, the ECOs have non-zero reflectivity, thus we may expect a non-zero Love number. To our expectation, it indeed turns out to have a non-trivial expression, whenever the reflectivity is non-zero, such that,
\begin{widetext}
\begin{equation}
\label{lonum}
k_{\ell m}(\omega)=\frac{(\ell-2)! (\ell+2)!}{(2\ell)! (2\ell+1)!}\times \textrm{Re}\Bigg[\frac{i\left(2M\omega\, r_{+}-am\right)}{2\left(r_{+}-r_{-}\right)} \left\{\frac{1-\mathcal{R}\,\gamma(\omega)}{1+\mathcal{R}\,\gamma(\omega)}\right\} \prod_{j=1}^{\ell}\left(j^{2}+4\frac{\left(2M\omega r_{+}-am\right)^{2}}{\left(r_{+}-r_{-}\right)^{2}}\right)\Bigg]~,
\end{equation}
\end{widetext}
where, $\mathcal{R}$ is the reflectivity of the boundary of the compact object. In particular, for a quantum black hole, $\mathcal{R}$ has a very specific expression, which ensures that the object absorbs only the characteristic frequencies and reflects all the other frequencies. The above expression for the tidal Love number depends on the quantity $\gamma(\omega)$, which is given by, 
\begin{align}
\label{gamma}
\gamma (\omega)\equiv \frac{\Gamma(\ell+3)\Gamma\left(-1-2ip\right) \Gamma\left(\ell+1+2ip\right)}{\Gamma(\ell-1)\Gamma\left(3+2ip\right) \Gamma\left(\ell+1-2ip\right)}~,
\end{align}
where, the frequency dependence enters through the quantity $p$, defined earlier as $p\equiv (am-2M\omega r_+)/(r_+-r_-)$. Observe that the tidal Love number $k_{\ell m}$ is a non-trivial function of the frequency $\omega$ and also depends on the reflectivity $\mathcal{R}$ of the surface of the ECO. As evident, the exact form of $k_{\ell m}$ is dependent on both the real and imaginary parts of the reflectivity, as well as on the real and imaginary parts of the complex and frequency-dependent quantity $\gamma(\omega)$. Thus, even if the reflectivity is a constant, the Love numbers will still vary with the perturbation frequency. Note that in the limit $p \to 0$, the quantity $\gamma(\omega)$ diverges, but the expression for the Love numbers remains finite. We can also express the Love numbers $k_{\ell m}$ in terms of the real and imaginary parts of $\gamma(\omega)$ as,
\begin{align}
k_{\ell m}(\omega)&=-p\left(\frac{(\ell-2)! (\ell+2)!}{(2\ell)! (2\ell+1)!} \right) \prod_{j=1}^{\ell}\left(j^{2}+4 p^{2}\right)
\nonumber
\\
&\qquad \times \left(\frac{\gamma_{r}(\omega) \mathcal{R}_{i}+\mathcal{R}_{r}\gamma_{i}(\omega)}{\left|1+\mathcal{R} \Gamma(\omega)\right|^{2}}\right)~. \label{loveexp} 
\end{align}
Above we have used the decomposition of the reflectivity as, $\mathcal{R}=\mathcal{R}_{r}+i\mathcal{R}_{i}$, and $\gamma(\omega)=\gamma_r(\omega)+i\gamma_i(\omega)$. The explicit form of $\gamma_r$ and $\gamma_i$ are given in \ref{AppendixA}.

In the limit of a classical black hole, the reflectivity $\mathcal{R}$ identically vanishes, and we observe that the tidal Love number vanishes as well, thereby recovering the results of \cite{Chia:2021}. On the other hand, in \ref{loveexp}, there is an overall factor of $(2M\omega r_{+}-am)$. Therefore, in the $\omega\rightarrow 0$ case, for a non-rotating compact object, we have $k_{\ell m}(\omega)\rightarrow 0$, irrespective of the nature of the compact object. This seems to contradict the non-zero static Love numbers obtained for compact objects with non-rotating geometries in \cite{Hinderer:2007mb,Cardoso:2017cfl}. This apparent discrepancy can be attributed to the existence of different branches of solutions of the hypergeometric differential equation, as discussed in \ref{AppendixB}. Thus, in the presence of non-zero reflectivity, the Love numbers are non-vanishing and frequency dependent. This is the main result of our work. We emphasize that this fact is sufficiently general and does not depend on the details of the nature of the compact object.

\section{Modeling the reflectivity of a quantum black hole}\label{refquant}

So far, our analysis was completely general and applies to any class of ECOs. In this section, as an illustration of the application of \ref{loveexp}, we consider the particular case of an area-quantized black hole. As described in \ref{area_quant}, the absorption of an area-quantized black hole takes place only at the characteristic frequencies, $\omega = \omega_{N,n}$, fixed by the quantization rule. These frequencies depend on the details of the quantization scheme, and can be found in \ref{uniform_area} for uniform area quantization, while \ref{subleading_area} provides the result for non-uniform area quantization. At these characteristic frequencies, the compact object behaves as a classical black hole, and hence the reflectivity will vanish, while for any other frequencies there will be a non-zero reflectivity. Therefore, the reflectivity of an area-quantized black hole may be modelled as a sum of Hann Window functions centered at the characteristic frequencies $\omega_{N,n}$, such that \cite{Datta:2021row},
\begin{align}
\mathcal{R}_{\rm q}=e^{\frac{i\pi M\omega}{2}}\bigg[1-\sum_n \mathfrak{P}\left(\omega-\omega_{N,n},D/2\right)\bigg]~. \label{reflectivity_exp}
\end{align}
Here, $\mathfrak{P}$ becomes unity at the characteristic frequencies $\omega_{N,n}$ (as we will see, some of these frequencies are indeed within the LIGO/VIRGO frequency band \cite{Agullo:2020hxe}) and hence, the reflectivity identically vanishes as it should. In addition, we have introduced a width $D$, which depends on the mass and spin of the black hole and is determined from the fitting formula used in \cite{Datta:2021row}\footnote{Note that, in \cite{Datta:2021row}, the width was denoted by $\Gamma$.}. Also, the phase of the reflectivity $\mathcal{R}_{\rm q}$ is expected to depend on the dimensionless combination $M\omega$. We emphasize that the expression in \ref{reflectivity_exp} is based on reasonable physical considerations, since a concrete microscopic computation of the reflectivity of a area-quantized black hole does not exist. Given this reflectivity, one can substitute the same in \ref{lonum} and using the expressions for $\gamma(\omega)$ from \ref{AppendixA}, the frequency-dependent tidal Love numbers can be computed. In the subsequent section, we will explore the consequences of our Love number expression for a reflecting compact object, in particular for a quantum black hole in the context of its observability in the present or future generations of the gravitational wave detectors. 
 
\begin{figure}[h!]
\includegraphics[width=\columnwidth]{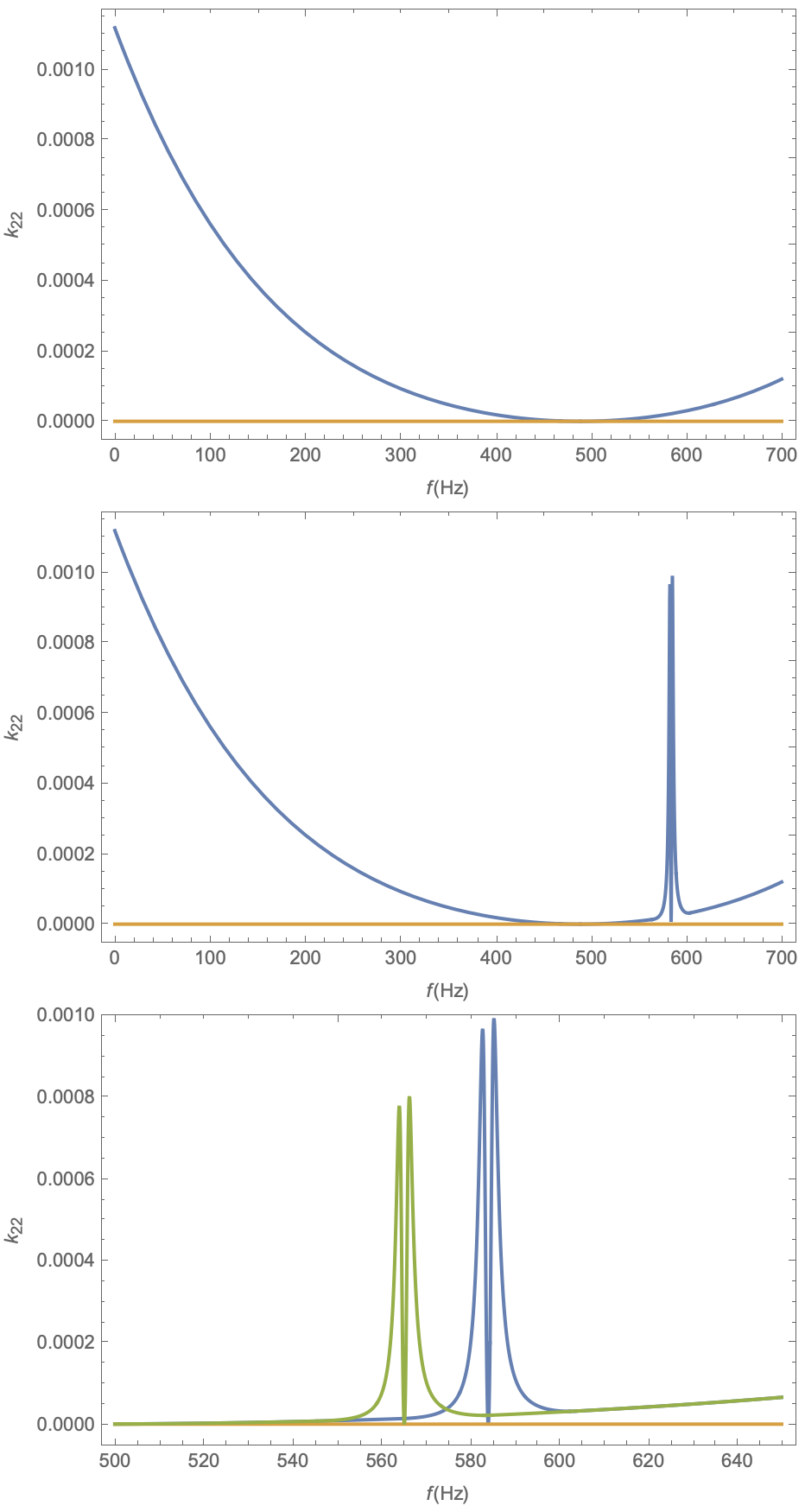}
\caption{The tidal Love number for the most dominant mode $k_{22}$ has been presented as a function of $\omega$ for various area-quantized black holes with mass $M=10M_\odot$ and $\chi=a/M=0.3$. The orange line is for a classical Kerr black hole, which predicts zero tidal Love number. The first two figures from the top describe uniform quantization with $\alpha=8\pi$ and $4\log 2$, respectively. The last figure compares uniform (blue) and non-uniform (green). Having $\alpha=4\log2$ and $C=-1.38,\nu=-0.01$. Note that the frequency where $P_+=0$ should be excluded from these curves because of the different branches of solutions, as discussed earlier.}
\label{plots_LIGO}
\end{figure}

\section{Implications for gravitational wave observations}\label{imp_gw}

Having derived the general expression for the tidal Love number, along with the reflectivity of an area-quantized black hole, we will now compute\footnote{Remember that the computation of the tidal Love numbers, following \ref{loveexp}, is based on the small frequency assumption, i.e., the expression is only applicable for those frequencies which satisfy $f \ll (1/2\pi)(c^{3}/GM)$. For an object of mass $10M_{\odot}$, it follows that $M\omega\sim\mathcal{O}(10^{-1})$, around $700\textrm{Hz}$.} the tidal Love number of the most dominant $(\ell=2,m=2)$ mode, for frequencies within the range of the LIGO/VIRGO collaboration. The tidal Love number $k_{22}$ contains information regarding the details of the compact object and the near horizon physics through the reflectivity. We will now demonstrate how one may extract this information from the observations using the area-quantized black holes. In this case, the details of the underlying quantum theory can be inferred through the parameter $\alpha$ for the uniform area quantization and the parameters $(\alpha,C,\nu)$ for non-uniform area quantization. As discussed in \ref{area_quant}, $\alpha$ is expected to lie somewhere between $8\pi$ and $4\log2$, so in our discussion, we will consider these two values. To illustrate the effects of non-uniform quantization, we have chosen parameters  $C = -1.38,\nu = -0.01$ so that there is no overlap between energy lines \cite{Chakravarti:2021jbv}. The corresponding results have been plotted in \ref{plots_LIGO}. 

As evident from \ref{plots_LIGO}, in the case of uniform area quantization, for $\alpha=8\pi$, the Love number is a non-trivial function of frequency and has no sharp features in the frequency range for which the calculations are valid. This is because, for this value of $\alpha$, there is no transition frequency within $700~\textrm{Hz}$. But when $\alpha=4\log 2$, we will see new features with width $D$ starting to show up in the spectrum of the Love numbers. This is because there are allowed transition frequencies within the frequency range of $f=700~\textrm{Hz}$. As expected, the Love number vanishes exactly at the characteristic frequencies. Further, the width $D$ for a $10M_\odot$ black hole is about 40 Hz and thus can be resolved by the LIGO/VIRGO collaboration. This means that by studying the sharp features in the love number spectrum, one can infer the details of the near horizon physics and also the value of the parameters associated with the quantum black hole. On the other hand, for non-uniform area quantization, the features identified for uniform quantization with a given $\alpha$ will be displaced from their expected position in frequency space, as demonstrated in \ref{plots_LIGO}. Therefore, the analysis of the spectrum of the tidal Love number can not only provide information about the quantum nature of black holes but will also tell us about the nature of the underlying quantum gravity through the specifics of the quantization mechanism. 

We can also predict what the future gravitational wave detectors, e.g., LISA, can say about the quantum nature of the black holes from the Love number observations. The analysis presented here will be more important for LISA than LIGO/VIRGO, as the effect of the tidal Love number arises in the waveform at the $\sim \mathcal{O}(2.5~\textrm{PN})$ order, which is a much higher order correction, given the smallness of the tidal Love number, as evident from \ref{plots_LIGO}. For LISA, with extreme-mass-ratio objects, the inspiral happens over a much longer duration; therefore, though a small effect, it can eventually build up during the prolonged inspiral, making it observable for LISA. For this purpose, we provide an estimate of the tidal Love numbers for a supermassive black hole. If we consider the mass of the supermassive black hole to be $M=10^{4}M_{\odot}$, the small frequency approximation will be valid till $f = 700~\textrm{mHz}$, well within the proposed sensitivity level of LISA. Thus the story will effectively repeat itself for LISA, but with better chances of being detected, due to the prolonged inspiral phase of the extreme-mass-ratio objects. The generalization of our results for quantum black holes of different origin \cite{Ashtekar:1997yu,Ashtekar:2000eq,Brustein:2020tpg,Brustein:2021bnw,Brustein:2021pof}, and calculation of their love numbers will follow an identical pattern.
\section{Conclusions}\label{section:conclusion}

Classical black holes have vanishing Love numbers and cannot be deformed by any external tidal perturbations. Thus any departure of the tidal Love numbers from zero would signify deviations from the classical black hole paradigm. We have demonstrated that any horizon-less ECO will have a frequency-dependent tidal Love number, arrived using gauge invariant formalism, which may lead to observational signatures in the gravitational wave spectrum. In order to probe such modifications in the tidal Love number, we consider the case of an area-quantized black hole. Intriguingly, the variation of the tidal Love number with frequency have distinct features, which can not only distinguish quantum black holes from ECOs, but can also distinguish between various models of quantum gravity. As an example, for an object of mass 10 $M_{\odot}$, if it is a quantum black hole, then the spectrum of the tidal Love number will demonstrate sharp features, with the Love number vanishing to zero for certain frequencies, in contrast with other models of the ECOs, and all of these features are within the frequency band of LIGO/VIRGO collaboration. Similarly, we can also distinguish between various models of quantized black holes. If such sharp features are observed in the frequency spectrum of the tidal Love number in the low-frequency regime, they will favour the non-uniform area-quantized black hole model. In particular, the absence of such sharp features will suggest that the compact object is not an area-quantized black hole, or, is a uniform area-quantized black hole with a larger values of $\alpha$. It would be interesting to compute the love number spectrum for other quantum black hole scenarios as well. Another interesting avenue would be to study the tidal absorption of the ECOs and hence of the area quantized black holes, whose relation to the well-known phenomenon of tidal heating is still unknown and remains an open question to ponder upon.  
\section*{Acknowledgements}

The authors are thankful to Rajes Ghosh for extensive discussion. S.N. and S.S. also thank IACS, Kolkata, for hospitality, where part of this work was carried out. Research of S.C. is funded by the INSPIRE Faculty fellowship from DST, Government of India (Reg. No. DST/INSPIRE/04/2018/000893) and by the Start-Up Research Grant from SERB, DST, Government of India (Reg. No. SRG/2020/000409). The research of S. S. is supported by the Department of Science and Technology, Government of India, under the SERB CRG Grant (No. CRG/2020/004562). Research of S.N is supported by the Prime Minister's Research Fellowship (ID-1701653), Government of India. 
\appendix
\labelformat{section}{Appendix #1} 
\labelformat{subsection}{Appendix #1}
\section{Explicit forms of the real and imaginary parts of $\gamma(\omega)$}\label{AppendixA}

We have provided the expression of the quantity $\gamma(\omega)$ in \ref{gamma} of the main text. However, the Love number computation depends explicitly on the real and imaginary parts of the same. Here we provide the explicit forms of the real and imaginary parts of $\gamma(\omega)$, as,
\begin{equation}
\gamma(\omega)=\gamma_r(\omega)+i\gamma_i(\omega)
\end{equation}
where the real part $\gamma_{r}$ and the imaginary part $\gamma_{i}$ have the following expressions (the frequency dependence enters through the definition of the quantity $p=(am-2M\omega r_+)/(r_+-r_-)$, as in earlier cases),
\begin{widetext}
\begin{align}
\gamma_r(\omega)&=\frac{(\ell-1) \ell(\ell+1)(\ell+2)}{4 p\left(16 p^{6}+24 p^{4}+9 p^{2}+1\right) \prod_{j=1}^{\ell}\left(j^{2}+4 p^{2}\right)} \left[p\left(4 p^{2}-5\right)\left(\Sigma_r^{2}- \Sigma_i^{2}\right)-2\left(8 p^{2}-1\right) \Sigma_r \Sigma_i\right]~,
\\
\gamma_i(\omega)&=\frac{(\ell-1) \ell(\ell+1)(\ell+2)}{4 p\left(16 p^{6}+24 p^{4}+9 p^{2}+1\right) \prod_{j=1}^{\ell}\left(j^{2}+4 p^{2}\right)}\left[2 p\left(4 p^{2}-5\right) \Sigma_r  \Sigma_i+\left(8 p^{2}-1\right)\left(\Sigma_r^{2}- \Sigma_i^{2}\right)\right]~,
\end{align}
where, 
\begin{align}
\Sigma_r&=\left\{\begin{array}{l}
            (2ip)^{\ell}+C_{2}\left(2 i p\right)^{\ell-2}+C_{4}\left(2 i p\right)^{\ell-4}+\cdots+C_{\ell-2}\left(2 i p\right)^{2}+\ell !,\quad\frac{\ell}{2} \in \mathbb{Z} 
            \\
            C_{1}\left(2 i p\right)^{\ell-1}+C_{3}\left(2 i p\right)^{\ell-3}+C_{5}\left(2 i p\right)^{\ell-5}+\cdots+C_{\ell-2}\left(2 i p\right)^{2}+\ell !,\quad\frac{\ell}{2} \notin \mathbb{Z}
            \end{array}\right.
            \\
\Sigma_i&=\left\{\begin{array}{l}
                    C_{1}\left(2 p\right)^{\ell-1} i^{\ell-2}+C_{3}\left(2 p\right)^{\ell-3} i^{\ell-4}+\cdots+C_{\ell-3}\left(2 p\right)^{3} i^{2}+C_{\ell-1}\left(2 p\right),\quad\frac{\ell}{2} \in \mathbb{Z} \\
                    \\
                    (2p)^{\ell}i^{\ell-1}+C_{2}\left(2 p\right)^{\ell-2} i^{\ell-3}+C_{4}\left(2 p\right)^{\ell-4} i^{\ell-5}+\cdots+C_{\ell-3}\left(2 p\right)^{3} i^{2}+C_{\ell-1}\left(2 p\right),\quad\frac{\ell}{2}\notin\mathbb{Z}
                    \end{array}\right.
\end{align}
The constants $C_{1}$, $C_{2}$, $C_{3}$ and so on, appearing in the above expression can be expressed as,
\begin{align}
C_1&=1+2+3+\cdots+(\ell-1)+\ell
\\
C_2&=1.2+2.3+1.3+3.4+1.4+\cdots+(\ell-2)(\ell-1)+(\ell-1)\ell
\\
C_3&=1.2.3+2.3.4+1.3.4+3.4.5+\cdots+(\ell-2)(\ell-1).1+(\ell-2)(\ell-1)\ell
\end{align}
\end{widetext}
The above expressions can be obtained by repeatedly using the properties of the Gamma function from Section 1.2 of \cite{eqbook}. We have used these results, while computing the tidal Love number of the area-quantized black holes. 

\section{Static limit of the dynamical Love numbers}\label{AppendixB}

For studying the static love numbers of a non-rotating configuration, we need to consider the differential equation \ref{radialdif} with $p=0$, which yields,
\begin{equation}
    \begin{aligned}
        &\frac{d^{2} \mathcal{R}}{d \xi^{2}}-\left[\frac{1}{\xi}+\frac{1}{(\xi+1)}\right] \frac{d \mathcal{R}}{d \xi}\\
        &\text{\hspace{3cm}}-\left[\frac{\ell(\ell+1)-2}{\xi(\xi+1)}\right] \mathcal{R}=0~.
    \end{aligned}
\end{equation}
The general solution to the above differential equation is a linear combination of the associated Legendre polynomials. These are given by,
 \begin{equation} \label{sol0}
    \begin{aligned}
          \mathcal{R}(\xi)&=A\xi(\xi+1)P_\ell^2(2\xi+1)+B\xi(\xi+1)Q_\ell^2(2\xi+1)\\
          &= \xi(\xi+1)\{A P_\ell^2(2\xi+1)+B Q_\ell^2(2\xi+1)\}\\
          &= \xi(\xi+1)H_0(2\xi+1)~.
    \end{aligned}
 \end{equation}
Where, in the last step, we have identified the object in curly brackets as $H_0(\xi)=A P_\ell^2(2\xi+1)+B Q_\ell^2(2\xi+1)$. One may notice that this $H_0$ is exactly the function that characterises the polar perturbations of the $g_{tt}$ component of the metric \cite{Macedo:2013qea,Hinderer:2007mb}. Thus, it follows that the Weyl scalar $\psi_{4}$ in the zero frequency limit, is simply given by $\psi_4\sim H_0(\frac{r}{M}-1)$. Thus, the tidal Love number in the zero frequency limit will be determined by the associated Legendre polynomials and will be identical to the ones presented in \cite{Macedo:2013qea}. Therefore, if one takes the zero frequency limit at the level of the differential equation the tidal Love number turns out to be consistent with earlier results. 

However, if we start from \ref{lonum} in the context of a non-rotating compact object and then take the zero frequency limit, we obtain vanishing tidal Love numbers. This apparent contradictory behaviour is hidden in the Hypergeometric differential equations, which have different branches, behaving differently in the zero frequency limit. To see precisely the cause of this behaviour, we can convert \ref{radialdif} into the more familiar form of a hypergeometric differential equation by using the following change of variable, $\mathcal{R}(\xi)=\xi^2(1+\xi)^2\mathcal{F}(-\xi)$, by which the \ref{radialdif} takes the form
\begin{equation}
\begin{aligned}
&-\xi(1+\xi)\frac{d^2\mathcal{F}(-\xi)}{d^2(-\xi)}+(3+2ip+6\xi)\frac{d\mathcal{F}(-\xi)}{d(-\xi)}\\
&\qquad\qquad +\left[\ell(\ell+1)-6\right]\mathcal{F}(-\xi)=0~.
\end{aligned}
\end{equation}
\\
Which is the Hypergeometric differential equation for $\mathcal{F}(-\xi)$ with the following choice of parameters: $a=3+\ell$, $b=2-\ell$ and $c=3+2ip$. It follows that the Hypergeometric differential equation has different branches of solutions based on whether or not $c$ is an integer. In the present scenario as well, $p=0$ and $p\neq0$ correspond to situations where c is an integer or not. Thus the solutions to \ref{radialdif} correspond to different branches depending on whether $p$ is zero or not. This demonstrates that the branch of the hypergeometric equation contributing in the dynamical case is different from the one contributing to the static case. As discussed before, we can get around this difficulty by treating the dynamical and the static case separately, reproducing the results of \cite{Hinderer:2007mb,Cardoso:2017cfl}. 


\bibliographystyle{plainnat}
\bibliographystyle{./utphys1}
\end{document}